# A Centralized Optimization Approach for Bidirectional PEV Impacts Analysis in a Commercial Building-Integrated Microgrid

Jubair Yusuf, A S M Jahid Hasan, Luis Fernando Enriquez-Contreras, Sadrul Ula

Department of Electrical and Computer Engineering

University of California Riverside

*Abstract*--**Building sector is the largest energy user in the United States. Conventional building energy studies mostly involve Heating, Ventilation, and Air Conditioning (HVAC), and lighting energy consumptions. Recent additions of solar Photovoltaics (PV) along with other Distributed Energy Resources (DER), particularly Plug-in Electric Vehicles (PEV) have added a new dimension to this problem and made it more complex. This paper presents an avant-garde framework for selecting the best charging/discharging level of PEV for a commercial building-integrated microgrid. A typical commercial building is used as a microgrid testbed incorporating all the DERs presented in a smart building. A Mixed Integer Linear Programming (MILP) problem is formulated to optimize the energy and demand cost associated with this building operation. The cost function is solved in conjunction with real data and modified to assess the bidirectional PEV impacts on the flexible building loads that are contributing factors in making energy usage decisions. Finally, the impacts of optimized DERs are investigated on a Distribution System (DS) to show the necessity of a holistic approach for selecting the suitable PEV strategies. The results show that bidirectional fast PEV activities can provide higher cost reduction and less voltage deviation in comparison to slow PEV activities.**

*Keywords*-- building energy cost reduction; DER & PEV integration; distribution system; G2V & V2G; microgrid; MILP optimization.

NOMENCLATURE

**Continuous Variable**

**Index and Set**

| | |
|---|---|
| $t$ | Time slot index |
| $T$ | Set of Total time |

**Decision Variables**

| | |
|---|---|
| $P_{grid}$ | Power from the grid to the building (kW) |
| $P_{ch,B}$ | BESS charging power (kW) |
| $P_{disch,B}$ | BESS discharging power (kW) |
| $P_{G2V}$ | Power collected from the grid to vehicle (kW) |
| $P_{V2G}$ | Power transferred from vehicle to grid (kW) |
| $SOC_B$ | State of Charge of BESS (kWh) |
| $SOC_{EV}$ | State of Charge of EV (kWh) |
| $P_{HVAC}$ | Power consumed by HVAC (kW) |
| $P_{lighting}$ | Power consumed by lighting (kW) |
| $\varphi$ | Lighting power consumed per area (kW/ft$^2$) |
| $T_{setpoint}$ | Setpoint temperature for building (degree C) |

**Input Parameters and Continuous Variables**

| | |
|---|---|
| $P_{PV}$ | Power (kW) available from the solar PV array |
| $T_{out}$ | Outside temperature (degree C) |
| $C_e$ | Price of energy from the grid ($/kWh) |
| $GHI_{PV}$ | Global Horizontal Irradiance (GHI) of PV array (kW/m$^2$) |

**Input Parameters and Constants**

| | |
|---|---|
| $\Delta t$ | Time interval |
| $\eta_{PV}$ | Solar PV array efficiency |
| $A_{PV}$ | Area (m$^2$) of the solar PV array |
| $\xi_{PV}$ | PV array parameter co-efficient |
| $T_a$ | Ambient temperature (degree C) |
| $SOC_{min,B}$ | Minimum SOC for BESS (kWh) |
| $SOC_{max,B}$ | Maximum SOC for BESS (kWh) |
| $\eta_{ch,B}$ | BESS charging efficiency |
| $\eta_{disch,B}$ | BESS discharging efficiency |
| $P_{ch,B\ max}$ | Maximum charging power for BESS (kW) |
| $P_{disch,B\ max}$ | Maximum discharging power for BESS (kW) |
| $SOC_{min,EV}$ | Minimum SOC for EV (kWh) |
| $SOC_{max,EV}$ | Maximum SOC for EV (kWh) |
| $\eta_{ch,EV}$ | EV charging efficiency |
| $\eta_{disch,EV}$ | EV discharging efficiency |
| $P_{G2V\ max}$ | Maximum charging power for EV (kW) |
| $P_{V2G\ max}$ | Maximum discharging power for EV (kW) |
| $\varphi_{min}$ | Minimum lighting power consumed per area (kW/ft$^2$) |
| $\varphi_{max}$ | Maximum lighting power consumed per area (kW/ft$^2$) |
| $A_{Building}$ | Area (ft$^2$) of the building |
| $\eta_{lighting}$ | Lighting efficiency |
| $T_{setpoint\ min}$ | Minimum setpoint temperature for the building |
| $T_{setpoint\ max}$ | Maximum setpoint temperature for the building |
| $P_{misc}$ | Power consumed by miscellaneous building load |
| $C_{d,B}$ | Cost associated with battery activities ($/kWh) |
| $C_{d,EV}$ | Cost associated with EV activities ($/kWh) |
| $C_{dep\_daily}$ | Daily depreciation cost ($/day) |
| $E_{dep\_daily}$ | Daily depreciation energy (kWh/day) |
| $K_{dep}$ | Global coefficient of depreciation |
| $E_{BESS}$ | BESS Stored Capacity (kWh) |
| $E_{EV}$ | EV Stored Capacity (kWh) |

**Binary Variables**

| | |
|---|---|
| $b_1$ | Charging decision binary variable for BESS |
| $d_1$ | Discharging decision binary variable for BESS |
| $e_1$ | Charging decision binary variable for EV |
| $e_2$ | Discharging decision binary variable for EV |

# 1. Introduction

*1.1 Motivation*

Buildings in the United States account for 40 percent of the total energy consumption and carbon-dioxide emission and approximately 74 percent of electrical energy is used in residential and commercial buildings [1]. On the other hand, the transportation sector is responsible for about 28 percent of the total energy consumption of which only 5 percent is from renewable energy. Fossil fuel still contributes to the largest share of total energy consumption. About 79 percent of domestic energy was produced from fossil fuels in 2018. U.S Energy Information Administration (EIA) projects 1,200 billion kWh energy consumption for Plug-in Electric Vehicle (PEV) with high penetration scenario by the next 20 years [2-3]. PEVs will help utilize renewable fuel for transportation but also become an integral part of smart homes and buildings by providing energy services to the building, e.g. Vehicle to Grid (V2G) power, demand management, etc. As they are being integrated into regular residential or commercial buildings, their relatively high electrical energy consumption pattern is a concern. Moreover, PEV being a mobile storage, their behavior is stochastic in comparison to regular stationary building connected Battery Energy Storage System (BESS).

According to U.S. EIA, recent buildings in the U.S. are more energy efficient in comparison to buildings built before 1980. The buildings in non- Organization for Economic Co-operation and Development (OECD) countries are not that energy efficient. But the recent proliferation of DER to the buildings makes the problem more complicated to handle. It's not only about having a smart thermostat that controls room temperature, but also a matter of co-scheduling other components that are available to make the best use of energy. State mandates and global efforts for slowing down climate change need more deployment of renewable and sustainable energy resources. Due to state mandates, California is one of the leading regions that have been experiencing a higher growth of DER installation in comparison to other states. It has more than 7,000 MW installed DER capacity and has a target to achieve more than 12,000 MW capacity by 2020. Figure 1 shows the DER installation scenario of selected states in the U.S. [4].

The growth of renewable and DER over the last decade has brought the attention of centralized optimization algorithms to both consumers and utilities. Higher DER additions have created the infamous duck curve in California, where increased solar production during the daytime reduces demand significantly but can create a need for higher generator ramp-up rates in the grid in the evening. Energy-efficient measures along with available DERs are so important to the consumers behind the meters due to the recent rapid increase of electrical energy cost. Unlike residential customers, electricity cost mostly consists of two components for the commercial customers. One is demand (kW) charge and another one is energy (kWh) charge. The consumers are being charged for the maximum peak that happened in a certain period along with the charge for total energy usage at that time. The peak demand charge is way higher than the regular energy charge. The maximum demand that happens during any 15-minute intervals is usually selected by the utility as peaks. For a sizeable commercial building-integrated microgrid, distributed energy resources (DERs) like Plug-in Electric Vehicles (PEVs) can cause higher peaks than the regular peaks created by the building loads. A level II PEV at 7 kW charging rate is often higher than the typical peak load and level III PEV or fast charging at 50 kW can result in 10 times higher peaks. The impacts of fast and slow charging on the cost optimization strategy of a building integrated microgrid needs to be thoroughly investigated. The overall fluctuation on other controllable loads along with the different charging levels has not been widely tested. Moreover, most of the problems focus on solving the energy cost optimization problem rather than minimizing both the demand and energy charges. There are multiple stakeholders for any building-integrated microgrid. While the optimization focuses on the consumer end, the results of the optimal scheduling need to be analyzed from the Distribution System Operator (DSO) perspective too. The optimal solution for customers behind the meter side may not result in the least voltage variation on the DSO side.

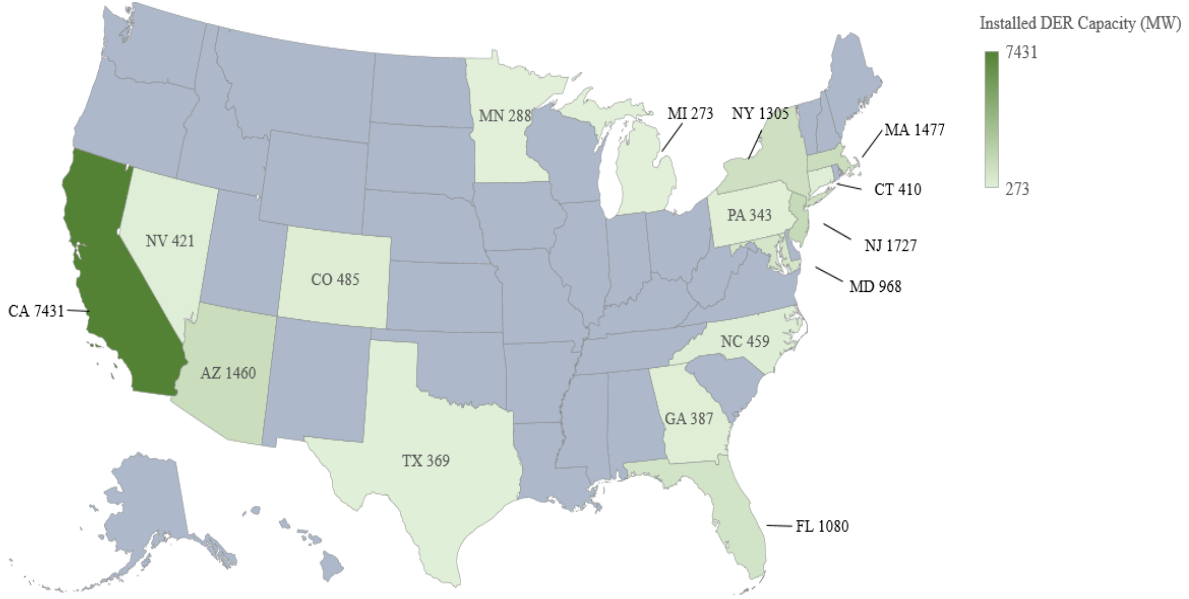

Fig 1: DER Installation (MW) Scenario in Selected U.S. States

State goals such as 100 percent renewable energy along with 5 million zero emission vehicles by 2045 and 2030, respectively, will add more complications to the overall energy optimization problem [5-6]. The upcoming wide PEV adoption needs mass scale workplace EV station establishment. Adoption of vehicle-to-grid protocols like ChadeMo or ISO 15118-20 has made it possible for the PEV owners to send power back to the grid to reduce the overall cost. This saving can be ensured in exchange for an added PEV degradation cost. Comprehensive model development is required to analyze all the issues corresponding to the different levels of bidirectional activities of PEVs. Extensive knowledge is needed to understand the flexibilities regarding the bidirectional PEV integration in a commercial building-integrated microgrid and deploy the best PEV framework.

*1.2 Related works*

Building-integrated microgrids energy use applications have given rise to many modeling and technical studies over the years. Demand response and energy reduction are the key features for any smart building and microgrid [7-8]. So most of the studies focus on optimal planning and operation of a building integrated microgrid irrespective of the presence of renewable and distributed energy resources. The research aspects of these studies can be divided into three categories in general. They are 1. modeling the building components and using the flexible building loads along with DERs (PV, BESS, PEV, etc.), 2. Optimal scheduling for optimization with different objectives, 3. Types of buildings (residential/commercial).

Most of the modeling and optimal solutions are derived for the HVAC and lighting components of the building in the earlier studies. Demand management strategies based on priority zones have been developed to reduce the peak demand which is the impacts of plug-in electric vehicle fast-charging in buildings. The strategy developed by energyplus modeling is more focused on buildings priority zones and fails to capture the dynamic properties of available DERs [9-10]. Hao et al. proposed a novel battery energy storage model for coordinating flexible building loads to provide end-user services [11]. The approach involves an aggregator-based co-scheduling platform for the betterment of both the consumer and power grid. In [12], Building flexibilities are used to reduce the distribution grid congestion using locational marginal prices.

In [13-20], a mixed-integer linear problem is formulated to schedule the energy consumption and minimize the total energy cost of a microgrid along with DERs. The model includes the plug-in loads but avoids the uncertainties associated with the intermittent energy sources and focuses on the tradeoff between carbon emission and energy cost. Different electricity pricing rates are applied to present various scenarios. The building parameters are modeled by US DOE energyplus software. Issues like BESS and PEV battery degradation have not been taken into consideration in these studies. Stochastic Dual Dynamic Programming is also used to solve the multi-stage stochastic problem that involved the bidirectional PEV approach and solar PV-based hybrid energy storage. But the issues like the level of charging/discharging and lower efficiency of PV-based

BESS are ignored during system optimization [21]. Yan et al. proposed a four-stage chance-constrained optimization and control algorithm for reducing the operational cost of a bidirectional EV charging station which ignored the modeling of the available building components [22]. A control algorithm is developed using available BESS, PEV, PV, and validated in a laboratory testbed to manage the peak and improve load factor in reference [23]. Operational cost optimization based on the occupancy profile is also examined in the literature [24-29].

Residential building microgrids are investigated by many researchers. In [30], a probabilistic home energy management approach is proposed for residential energy optimization and customer energy cost reduction with renewable sources. In [31], the combination of the thermostat and home energy management system is used for demand response in smart grid applications along with V2G. An optimal charging strategy for residential applications is proposed to shave the peak and reduce the energy cost [32]. In [33], a price-based demand response model is proposed to optimize EV charging and quantify the flexibility of EV charging. An optimization approach is proposed for multi-energy microgrid systems with the vehicle to grid capability and the optimal capacity is estimated too [34-35]. Igualada et al proposed a MILP for residential microgrid energy optimization where BESS degradation is not considered [36]. A smart energy management system is presented for a residential building with DERs and combined cooling and heating power systems [37]. All the studies are mainly focused on the temperature modeling of thermal load and use level I and level II charging during any EV integration. The willingness of EV activities of the PEV owners and the systems degradation costs are also not considered.

The usefulness of second-life EV batteries has been analyzed in terms of commercial building microgrid application and showed that it can be useful if used with other stationary batteries [38]. The energy cost in commercial buildings is optimized by the integrated control of dynamic facades [39], and HVAC [40]. In [41-42], A dynamic cost-effective solution for the participation of customers with DERs in grid transactions called DER-CAM is proposed. EV charging strategies [43] and EV owners' perspectives [44] are analyzed for an office building microgrid. Energy Management System (EMS) strategies are developed for a commercial [45-46] and an aggregated [47] building in the presence of EVs. In [48-51], the EMS is proposed for the campus prosumers to participate in the demand response programs. These lack the analysis of EV impacts on the nanogrid. The sensitivity of building loads on the cost optimization strategy along with the PEV and the energy flow of the PEV has been analyzed in references [52-53].

*1.3 Summary of Contributions*

All the relevant works discussed above lack a rigorous analysis of different levels of bidirectional PEV impacts on a commercial building-integrated microgrid. One can raise a number of questions regarding PEV integration: 1. How will the fast or slow PEV charging/discharging impact any grid-connected commercial building prosumer? 2. How will it interact with other controllable loads present in the building? 3. How will it impact the distribution grid while optimizing the savings of the customers behind the meter? 4. What happens to the bidirectional strategy when demand charge is integrated? 5. Which one is the best charging level for optimal cost and how much greater savings are possible in comparison to other charging/discharging levels? This paper tries to find the answers to all these questions and attempts to bridge this gap in the literature by offering an optimal framework for the best PEV charging/discharging level selection. The novel contributions of this paper can be summarized as follows:

1. Investigating PEV fast and slow charging/discharging impacts on the optimal scheduling problem of a commercial building-integrated microgrid.

2. Providing a robust and computationally efficient Mixed Linear Integer Programming solution to reduce the total cost. The degradation costs associated with the discharging of BESS and PEV are also integrated into the model in addition to the peak demand cost.

3. Helping the DSO in PEV Charging Station (CS) selection based on the optimal offline solution impacts and the sensitivity analyses between all the components present in a grid-connected commercial building-integrated microgrid.

4. Introducing a comprehensive framework for determining the best PEV charging/discharging level.

*1.4 Paper Organization*

This article is organized as follows: Section 2 describes the system and provides the modeling of the components, Section 3 discusses problem formulation and constraints, Section 4 includes the simulation results, and Section 5 concludes the paper with the discussion of future research directions.

**2. System description & modeling**

*2.1. System Details*

The testbed building is located at the College of Engineering – Center for Environmental Research & Technology (CE-CERT) at the University of California Riverside. This building has offices for research faculty, staff, students along with administrative offices and conference rooms. It is 21,352 sqft in size and serves as a distribution level microgrid for the existing utility feeder. It is equipped with 180 kW solar PV, a 500 kWh/100 kW BESS and, five EV charging stations. Four of the chargers are for level II charging (6- 7.2 kW) while the fifth one is a level III charger (50 kW). To perform a vehicle to grid (V2G) operation, there is a specialized bidirectional fast EV charger. EVs can send power back to the grid through this charger in addition to the usual charging mode of operation. This building consists of an open working space which means that there are no different zones. The pattern of energy use of this building is similar to a regular office building and is usually unoccupied during weekends and holidays. As energy-efficient measures are more needed when the building is occupied, a weekday has been selected to co-ordinate various loads and optimize the overall system. Figure 2 is showing the schematic of the system and available DERs.

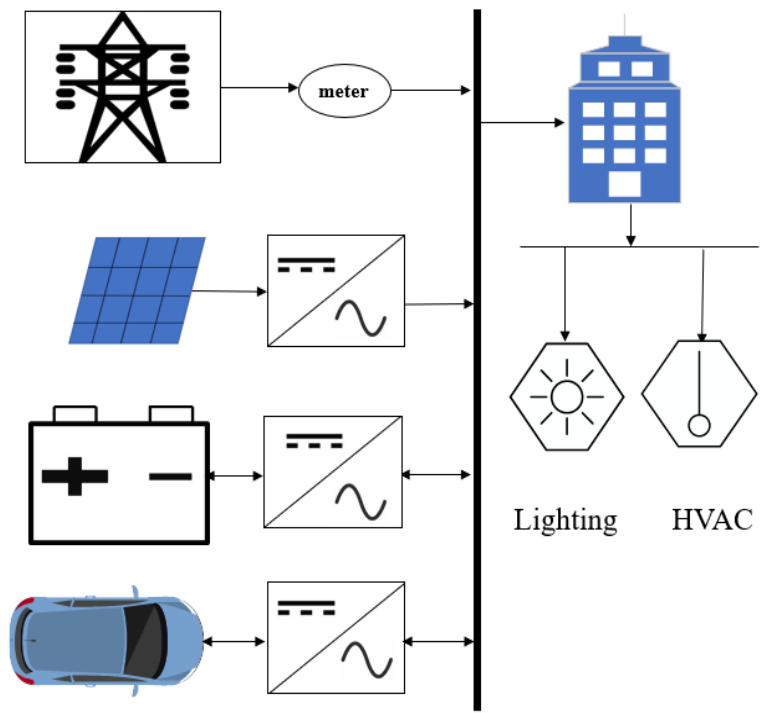

Fig 2: Building Schematic with behind the meter DER Components

*2.2. PV model*

There are different steady and dynamic state models available for estimating generated solar power. Time-series statistical models such as Autoregressive Integrated Moving Average (ARIMA), Seasonal Autoregressive Integrated Moving Average with Exogenous Variables (SARIMAX) are being used to predict solar power based on the availability of historical solar information. Recently, deep learning models are being used to predict time series data, but they are computationally expensive. The solar PV steady-state model based on temperature and irradiance provides the most accurate among all the available models and hence is widely used [54-55]. Equation (1) is used to estimate the possible solar production from this building's 180 kW solar PV. Global Horizontal Irradiance is used as it includes all the factors involved with estimating solar generation such as Direct Normal Irradiance (DNI), Diffuse Horizontal Irradiance (DHI), and ground reflected radiation. The irradiance and temperature are forecasted and collected by the weather station.

$$P_{PV}(t) = \eta_{PV} \times A_{PV} \times GHI_{PV}(t) \times \{1 - \xi_{PV}(T_{out}(t) - T_a)\} \text{ for } \forall t \in T \quad (1)$$

*2.3. BESS modeling*

Li-ion batteries are the most common batteries that are used for stationary Battery Energy Storage systems. Recently, flow batteries are also getting more attention. However, their inadequacy of maintaining enough State of Charge (SOC) is making them impractical for real applications. The system used in this study has a Li-ion battery as a stationary BESS whose equations are used for modeling in this study. Equation (2) is used to calculate the SOC of BESS. The constraint given in (3) ensures that the BESS charging rate always stays between the maximum and minimum charging limits whereas the discharging rate is regulated by (4). BESS is not allowed to do the charging and discharging simultaneously. This is prohibited by equation (5). Equation (6) states that both decision variables are binary. The BESS SOC limit is maintained by (7). Even though the available BESS capacity is significantly higher, the battery specifications considered suitable for this building's load profile are 150 kWh/50 kW rating with a 95 percent charging/discharging efficiency. The initial stored energy is assumed to be 105 kWh which is 70 percent of the total energy available. The minimum and maximum SOC are assumed 40 and 100 percent, respectively. The minimum SOC level may be lower, but 40 percent SOC is selected to have enough available stored energy in case of grid power failure.

$$SOC_B(t) = SOC_B(t-1) + \left(\eta_{ch,B} \times P_{ch,B}(t) - P_{disch,B}(t)/\eta_{disch,B}\right)/E_{BESS} \times \Delta t \quad \text{for } \forall t \in T \quad (2)$$

$$0 \leq P_{ch,B}(t) \leq b_1(t) \times P_{ch,B\ max} \qquad \text{for } \forall t \in T \quad (3)$$

$$0 \leq P_{disch,B}(t) \leq d_1(t) \times P_{disch,B\ max} \qquad \text{for } \forall t \in T \quad (4)$$

$$b_1(t) + d_1(t) = 1 \qquad \text{for } \forall t \in T \quad (5)$$

$$b_1(t), d_1(t) \in \{0,1\} \qquad \text{for } \forall t \in T \quad (6)$$

$$SOC_{min,B} \leq SOC_B(t) \leq SOC_{max,B} \qquad \text{for } \forall t \in T \quad (7)$$

*2.4. PEV modeling*

There are two charging protocols currently available for fast charging. One is the Combined Charging System (CCS) and another is CHAdeMO (Charge De Move). In this study, the Nissan Leaf available for V2G research has only the CHAdeMO protocol. Recent Teslas are not capable of V2G charging. For this study, Tesla battery capacity and charging rates are used to estimate the effects of level III charging on the overall optimization problem. Both Nissan Leaf E plus and Tesla X specifications are used for simulation purposes. The first one has a 64 kWh capacity with a power rating of 7 kW. The latter one has an energy storage capacity of 100 kWh and the maximum rate of charging or discharging is 50 kW. Equations and inequalities 8 through 13 represent the PEV battery in the model. As the PEVs have Li-ion BESS installed in their system, similar equations as in the BESS modeling are used to represent the PEV. Equation (8) measures the stored energy at each time instant t. Inequalities (9) and (10) help to control the maximum rate of charging and discharging for PEV and these rates vary with the level of charging/discharging chosen for PEV. Equations (11) and (12) ensure that PEV cannot try to perform G2V and V2G functions at the same time. The inequality (13) represents that the decision variable SOC of the PEV will always stay within the given bound. The focus of this study is finding out the optimal cost savings possible for a certain period while a PEV is available. For this testbed, there are always two PEVs present and one is available for bidirectional activities. Moreover, to address the uncertainties associated with PEV, it is modeled in such a way that it always has enough SOC left to finish 4 to 7 average round trips based on the PEV models. As the National Household Travel Survey (NHTS) in 2017 shows that about 58 percent of households in the U.S have more than one vehicle and 77 percent of trips are below 11 miles [56]. It neutralizes the range anxiety for any PEV user.

$$SOC_{EV}(t) = SOC_{EV}(t-1) + \left(\eta_{ch,EV} \times P_{G2V}(t) - P_{V2G}(t)/\eta_{disch,EV}\right)/E_{EV} \times \Delta t \text{ for } \forall t \in T \quad (8)$$

$$0 \leq P_{G2V}(t) \leq e_1(t) \times P_{G2V\,max} \qquad \text{for } \forall t \in T \quad (9)$$

$$0 \leq P_{V2G}(t) \leq e_2(t) \times P_{V2G\,max} \qquad \text{for } \forall t \in T \quad (10)$$

$$e_1(t) + e_2(t) = 1 \qquad \text{for } \forall t \in T \quad (11)$$

$$e_1(t), e_2(t) \in \{0,1\} \qquad \text{for } \forall t \in T \quad (12)$$

$$SOC_{min,EV} \leq SOC_{EV}(t) \leq SOC_{max,EV} \qquad \text{for } \forall t \in T \quad (13)$$

*2.5. HVAC modeling*

HVAC is responsible for typically the highest energy consumption in any building and is directly related to people's occupancy and behavior. Thermal load modeling for HVAC is popular and it depends on building design, air handling unit, chillers, and other components required to ensure thermal comfort for the occupants of the building [57]. Different building design models such as Energyplus, eQuest are widely used to do HVAC modeling for any building before construction based on assumptions. The assumptions are made based on the building size, climate region, and usage of the building. As the building already exists here, real electrical power consumption by HVAC units presented in the building is used for HVAC modeling to represent the actual scenario. A smart power analyzer Fluke meter is used to collect the electric power consumption by HVAC units for multiple days. HVAC power consumption is acquiescent with room temperature. The outdoor temperature is used as a reference temperature to find the correlation between the HVAC setpoint and power consumption. Later on, the Matlab curve fitting tool is used to fit a curve and generate the equation for HVAC power consumption with the available power and temperature data.

$$P_{HVAC}(t) = -0.2186 \times \left(T_{setpoint}(t) - T_{out}(t)\right) + 5.63 \qquad \text{for } \forall t \in T \quad (14)$$

*2.6. Lighting modeling*

Consumption of lighting highly depends on the application of the building. The light intensity of any area can be measured in lux. The recommended lux for the warehouse and work area are 0.1 and 0.15 kW/m2 respectively [58]. Equation 15 has been used to estimate the lighting power of the building and light intensity has been controlled by equation (16).

$$P_{lighting}(t) = (0.0929 \times \varphi(t) \times A_{Building})/\eta_{lighting} \quad \text{for } \forall t \in T \quad (15)$$

$$\varphi_{min} \leq \varphi(t) \leq \varphi_{max} \qquad \text{for } \forall t \in T \quad (16)$$

*2.7. Temperature comfort Modeling*

Thermal comfort is highly dependent on the thermal load, metabolic rate, and clothing insulation of people. Typical comfortable room temperature is used as a basis for the temperature comfort modeling here. Equation (17) is deployed to model the temperature comfortable for the building.

$$T_{setpoint\ min} \leq T_{setpoint}(t) \leq T_{setpoint\ max} \qquad \text{for } \forall t \in T \quad (17)$$

*2.8. Energy price modeling*

Three energy rates are used to analyze the pricing impacts on the optimization framework. The rate used for the base case is from an Investor-Owned Utility (IOU), Southern California Edison (SCE). It is the highest energy rate of all three utility rates. It is assumed that the electric vehicle is plugged on from 9 am-9 pm. As it is a Time of Use (TOU) energy rate for SCE, it charges consumers more between 4-9 pm. This scenario is highly different in other states where solar is not abundant. The maximum peak demand charge is 3.83 $/kW. Table 1 summarizes the energy cost for this period.

**Table 1**
TOU Based Energy Charge

| Time | IOU Energy Charge ($/kWh) |
|---|---|
| 9 am – 4 pm | 0.22 |
| 4 pm – 9 pm | 0.41 |

*2.9. Degradation cost modeling*

The depreciation cost for any li-ion battery is dynamic and depends on several factors such as daily usage, temperature, operation scenarios. The depreciation cost can be modeled using the equation (18) and it can vary from 0.08-0.13 $/kWh [59].

$$C_d = \frac{C_{dep\_daily}}{E_{dep\_daily}/K_{dep}} \quad (18)$$

*2.10. Power balance modeling*

Equation (19) describes the power balance equation for different energy providers. The sum of power provided by the grid to the building, solar PV, BESS, and EV will be equal to the power required for BESS charging, EV charging, HVAC, lighting, and miscellaneous load for the building.

$$P_{grid}(t) + P_{PV}(t) + P_{disch,B}(t) + P_{V2G}(t) = P_{ch,B}(t) + P_{G2V}(t) + P_{HVAC}(t) + P_{lighting}(t) + P_{misc}(t) \text{ for } \forall t \in T \quad (19)$$

## 3. Problem formulation and constraints

### 3.1. Objective function

The objective functions of the optimization problem are stated by (20) and (21). The first equation is needed to be solved for minimizing the energy cost and degradation cost resulted from BESS and PEV activities and the second equation also integrates the demand cost.

Objective 1:

$$\text{Minimize } \sum_{t=1}^{T} \left\{ \left(P_{grid}(t) \times \Delta t \times C_e(t)\right) + \left(\eta_{ch,B} \times P_{ch,B}(t) + P_{disch,B}(t)/\eta_{disch,B}\right) \times C_{d,B}(t) \times \Delta t + \left(\eta_{ch,EV} \times P_{G2V}(t) + P_{V2G}(t)/\eta_{disch,EV}\right) \times C_{d,EV}(t) \times \Delta t \right\} \quad (20)$$

Objective 2:

$$\text{Minimize } \left[ \max P_{grid} \times C_d + \sum_{t=1}^{T} \left\{ \left(P_{grid}(t) \times \Delta t \times C_e(t)\right) + \left(\eta_{ch,B} \times P_{ch,B}(t) + P_{disch,B}(t)/\eta_{disch,B}\right) \times C_{w,B}(t) \times \Delta t + \left(\eta_{ch,EV} \times P_{G2V}(t) + P_{V2G}(t)/\eta_{disch,EV}\right) \times C_{w,EV}(t) \times \Delta t \right\} \right] \quad (21)$$

### 3.2. Constraints

All constraints for the formulated optimization problem are described by equations (1)-(19).

*3.3. Optimization*

The objective functions and all the equations used as constraints are linear. Both continuous and binary variables exist in the sets of constraints' variables. Therefore, this optimization problem is a MILP problem. Gurobi Python [60] environment has been used to model and solve this optimization problem. The workstation used to solve it is of core-i7 with 16 GB RAM.

**4. Simulation results & Discussions**

*4.1. Base case*

The base case used to solve this optimization problem considers the availability of a regular good sunny day for maximum solar generation. Maximum thermal comfort is induced to make the temperature level varying in a very short range. The plug-in load is considered constant for this building. As battery degradation cost is considered for both BESS and PEV, so higher energy charges such as the California IOU energy rate are used to figure out the actual impacts of all the controllable and non-controllable loads and sources available. The specifications of all the base case variables are noted in Table 2.

**Table 2**

Base Case Description

| **Base Case** | |
|---|---|
| Solar PV | Sunny Day |
| BESS | 150 kWh/50 kW |
| PEV | 64 kWh/7 kW |
| Lighting | 50% of Total Building Area |
| Temperature | 24°C≤T≤26°C |
| Energy Price | Investor-Owned Utility |
| Miscellaneous | Constant |

*4.2. Slow and fast G2V/V2G impacts on the base case*

Solar power was mostly available from 9-3 pm for the given day. During the base case scenario, PEV operates in charging mode to reach the highest level of SOC while solar is abundant. Later, it shifts to discharging state and discharges power at a constant rate of 7 kW to reduce the total cost associated with G2V and V2G activities. This results in a sharp declination of PEV SOC and maintains the desired final SOC level at the end of the on-peak period. The lower capacity of PEV is the reason behind the constant discharging operation during on-peak periods. BESS acts in a similar way and discharges during off-peak periods. Figure 3 shows the impacts of all charging/discharging scenarios for PEV to optimize the energy cost.

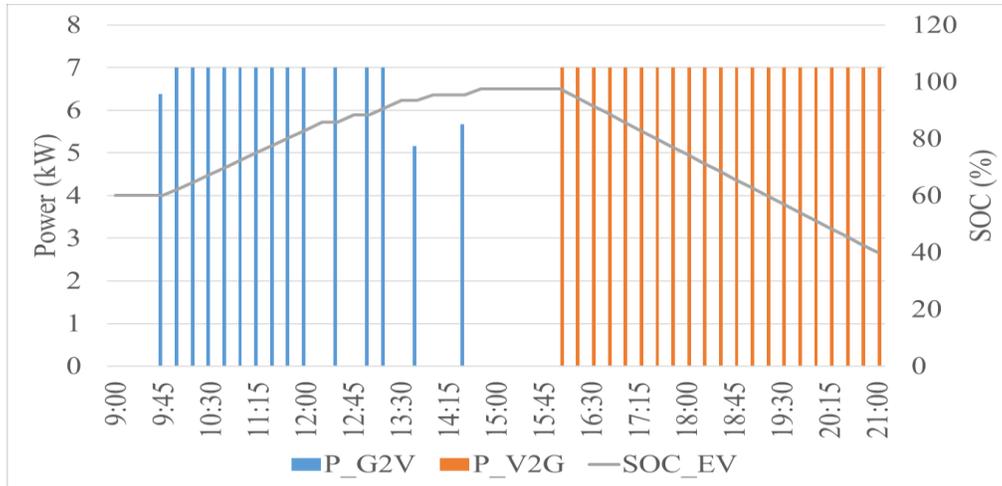

Fig 3: Base Case with Nissan Leaf E Plus: EV charging profile along with SOC

In the case of fast charging/discharging impacts analysis, the available EV is capable of DC fast charging with 480 V three-phase ac power connection and up to 50 kW G2V/V2G rate. Both EV and battery charging/discharging take place at the same time in the case of level II charging. In contrast, the PEV and battery charging/discharging activities do not happen simultaneously in this case. If the charging/discharging events take place less, the battery life for BESS and PEV will sustain more. So, the charging and discharging events happen for BESS when EV is not in action and vice versa. The G2V and V2G profiles for PEV along with relevant SOC and solar generation are shown in figure 4. Level III EV activities provide the higher cost-benefit for both objectives shown in Table 3. As no constraint is imposed upon the maximum demand that can be provided by the grid, so the optimal scheduling is the same for both objectives. The rest of the sensitivity analyses discussed in the latter sections are based on the first objective.

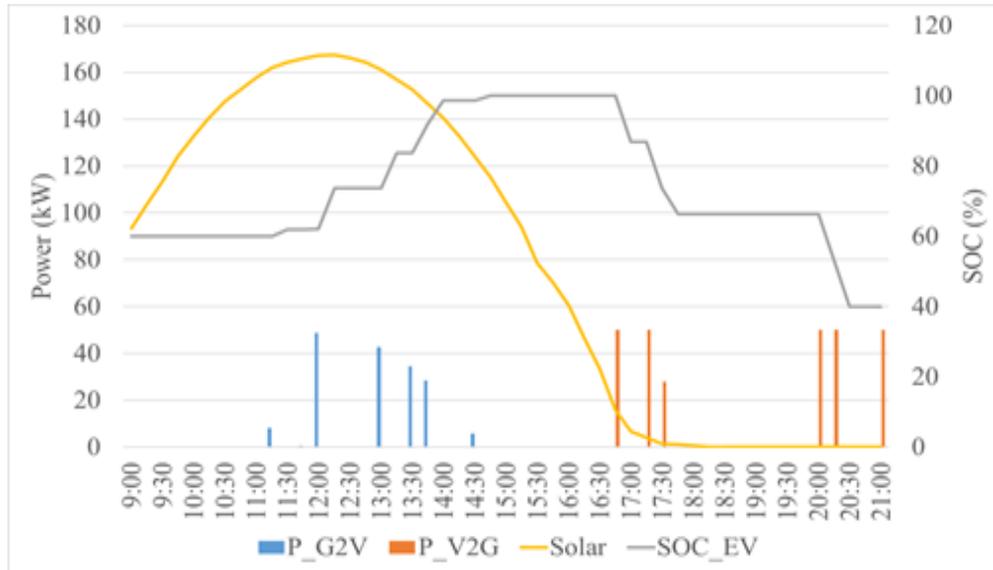

Fig 4: Base Case with Tesla X: Solar and EV charging profile along with SOC

**Table 3**

Base Case Daily Cost Comparison

| Type | Objective 1 | Objective 2 | Percentage of Savings (%) (Objective 1) | Percentage of Savings (%) (Objective 2) |
|---|---|---|---|---|
| Heuristic Solution | 273.34 | 457.18 | – | – |
| Nissan Leaf: Level II | 217.27 | 401.112 | 20.5 | 12 |
| Tesla X: Level III | 210.49 | 394.33 | 23 | 14 |

*4.3. Effects of lighting variation*

Lighting is one of the highest energy consumer loads present in buildings. To analyze the lighting impacts on the optimization strategy, lighting intensity is varied within a permissible limit. Advanced control systems help to vary the lighting of a place depending on its occupancy. Occupancy sensors and motion sensors are the most common control types of equipment for lighting variation. Centralized building automation control algorithms can include lighting control along with control of HVAC and other DER resources. The effects on the maximum peak and energy cost due to lighting variation are studied and documented. Level III charging/discharging capable electric vehicles always results in achieving minimum energy cost. Higher battery capacity and a higher rate of charging/discharging offer the best opportunity to charge/discharge at the right time. As expected, energy cost increases when the lighting level increases. In the case of a slow G2V/V2G level, for an increase of 40 to 60 percent lighting level, the maximum peak occurred with a 43 percent increase. About 59 percent of energy cost increments happen because of the high lighting level which is even higher than the peak demand increment.

On the other hand, a higher charging rate makes the maximum peak occur while level III EV is operational. Though the change in maximum peak (40%) is lower in comparison to level II activities, the difference in energy cost increment is higher than that of level II (63%). Figure 5 shows the impacts of lighting variation on cost optimization strategy and maximum peak.

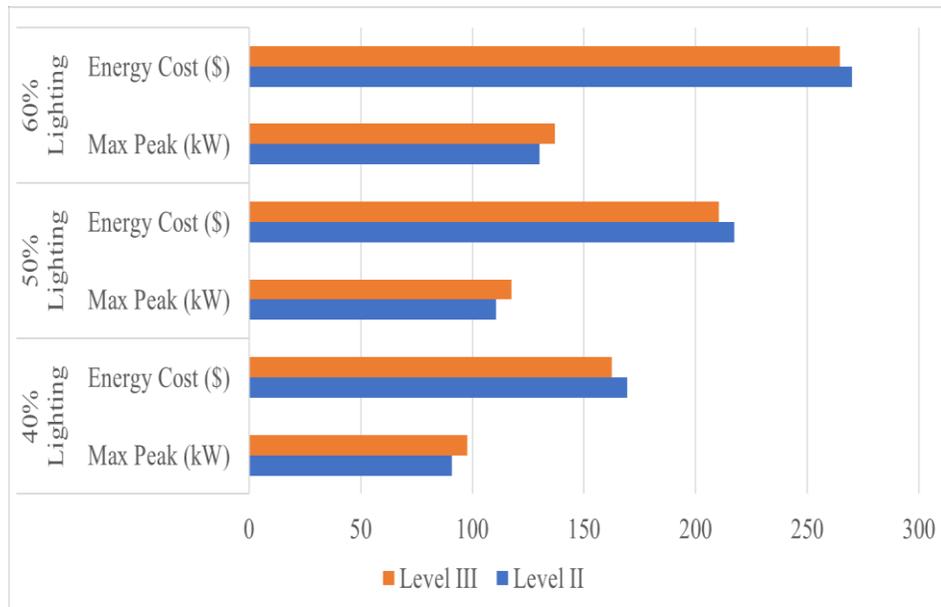

Fig 5: Impacts on Energy Cost and Maximum Peak for Lighting Variation

Figure 6 is showing the charging/discharging power of BESS and PEV along with their SOC for lighting level variation with fast charging. While a lower lighting level is selected for optimization, both BESS and PEV can work one at a time and provide maximum and longer energy support to the building. While a higher lighting level (60%) is required, then both BESS and PEV can not be fully charged during high solar availability. They are also incapable of providing energy during on-peak hours for a longer period.

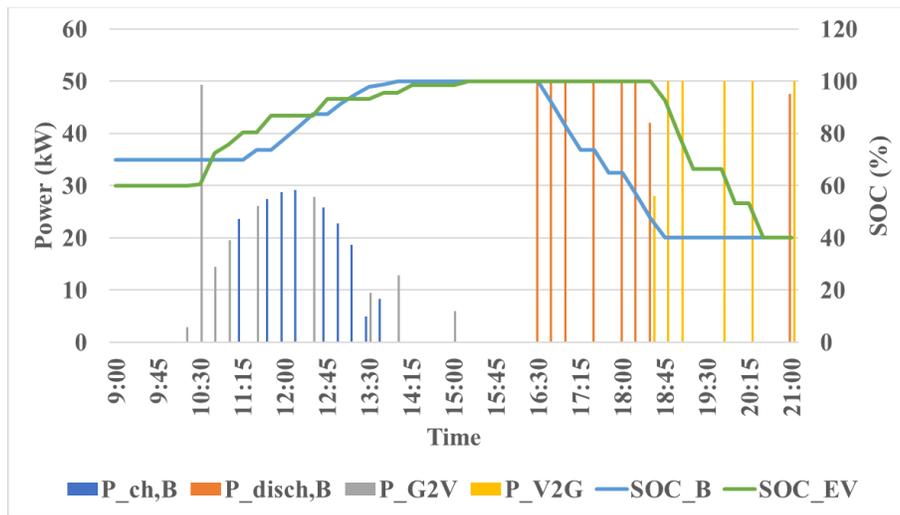

Fig 6: Base Case with Level III Charging and 40% Lighting: BESS & PEV Power and SOC Profile

*4.4. Effects of electricity price*

The pricing structure is a large motivating factor in the cost optimization strategy for any building integrated with DERs. If the price varies a little or does not vary at all over time, then the evaluation of decision variables gets more sophisticated. The pricing of energy largely depends on the size and average electricity usage by the building occupants. Typically, users of most small or medium-size buildings do not pay any demand (kW) charge. However, their energy (kWh) charges are significantly higher making energy cost optimization more important. Large energy consumers usually are on Time of Use (TOU) rates and must pay various demand charges in addition to the typically lower energy charges.

Two pricing structures from local Municipal Utility are used to test the impacts of pricing on our comprehensive optimization solution. The energy price per kWh is lower than the BESS and EV degradation cost on the first case and the on-peak energy cost is slightly higher than the off-peak energy charge. Later on, a flat energy charge is also used to find out the significance of optimal strategy for all the controllable resources available. The different energy charges of the public municipal utility are shown in Table 4.

**Table 4**

Public Municipal Utility Charge

| Type | Time | Energy Charge ($/kWh) |
|---|---|---|
| TOU | 9 am – 4 pm | 0.0874 |
| | 4 pm – 9 pm | 0.1079 |
| Flat Rate | 9 am – 9 pm | 0.1684 |

When this TOU rate is used for the optimization model, the BESS and PEV don't charge or discharge at all. As their degradation cost per kWh for any charging or discharging activities is higher than the energy cost, they don't take part in any energy minimization from the grid. Only other available controllable loads such as HVAC and lighting help to minimize the overall energy cost. On the other hand, with a flat energy price higher than the degradation cost, BESS and PEV do not take part in charging or discharging because of a lack of time-varying pricing opportunities. Their only usages are for managing the demand and intermittent solar production. Flat pricing provides 25.6% and 26.0% cost savings for level II and III charging, respectively.

*4.5. Effects of temperature variation*

Keeping tight constraints on temperature settings allow very comfortable and stable indoor temperatures. But it limits the opportunity for varying HVAC loads to contribute towards operational cost reduction. In this scenario, relying on the thermal mass of the building, the temperature is varied to a larger extent during the low occupancy period. The optimization problem is solved with a lower bound of 22 degrees and an upper bound of 28 degrees Celsius (25 +/- 3 °C). This larger temperature variation gives a slightly better economic benefit (20.9%-23.4%) compared to the base case. Raising the temperature limits further especially in extreme cases can result in more economic benefits, for example during critical grid events such as Flex Alert in California [61]. Flex alert is issued by the local utility when consumers automatically respond to reduce their energy usage and help the grid operators to prevent the dearth of power supply.

Figure 7 is showing the comparison of HVAC power consumption with the base case for level II charging activities. HVAC power consumption gets lower throughout the day for a wider range of temperature settings as expected during slow charging. It follows the same trend during fast charging activity except in one instant.

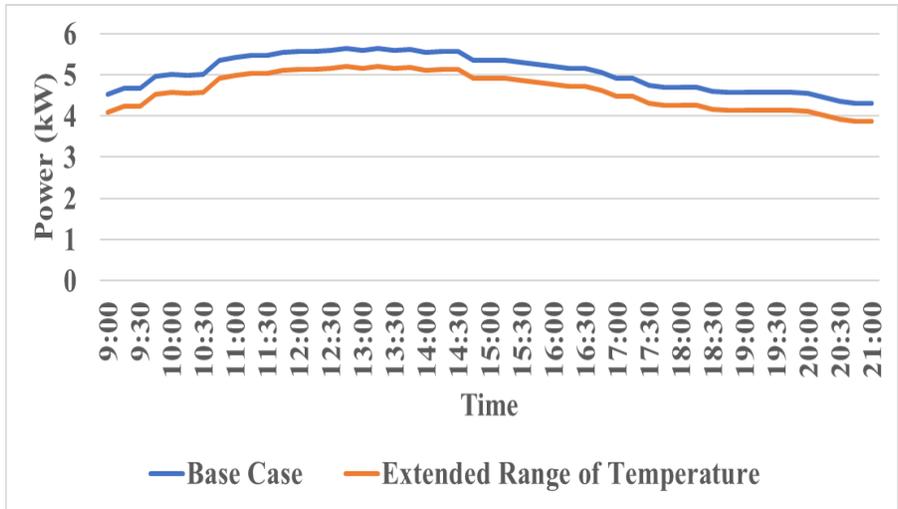

Fig 7: HVAC power consumption with Level II Activity: Base Case & Extended Range of Temperature

Figure 8 is showing the charging and discharging profile for BESS and PEV in case of extended variation in temperature and fast PEV activity. Their operations are steadier in comparison to base case temperature limits.

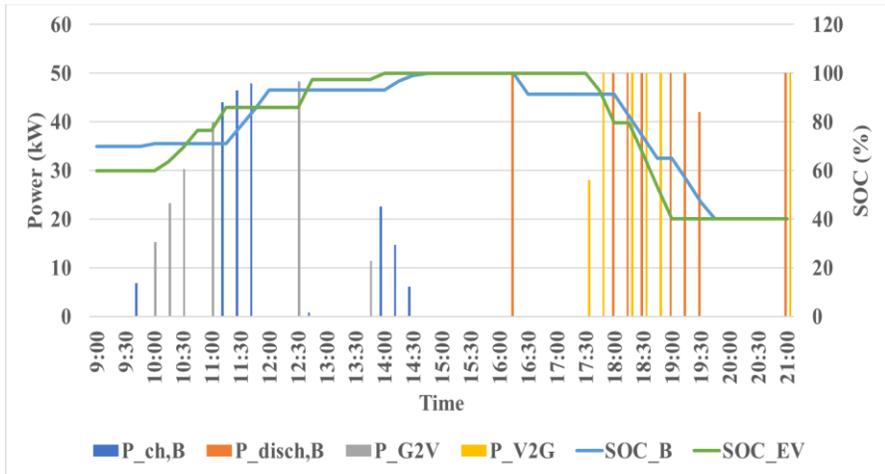

Fig 8: Temperature Variation with Extended Range & Level III Activity: BESS & PEV Power and SOC Profile

*4.6. Cloudy day impacts*

The BESS and PEV must get fully charged before the more expensive on-peak period on a sunny day so that they can discharge while needed. But when there is not enough solar during the daytime on a cloudy day, the situation gets worse. To evaluate the impacts of a cloudy day on the regular operation of the building with DERs, a cloudy day temperature, and solar irradiance profiles are used. The amount of savings possible for a cloudy day is much lower than a regular sunny day. A maximum of 6.1 percent energy cost reduction is possible for a cloudy day. A level III charger is better again in the case of energy cost optimization compared to a level II and provides 1.2 percent more savings.

Figures 9 and 10 are showing the net power imported from the grid for a typically sunny and cloudy day while the base case is considered. Cloudy day results in a 4.38 percent increase in peak during level II activities compared to a sunny day. On the other hand, a cloudy day results in a 1.75 percent decrease in peak demand for level III activities. The difference in peaks is due to the individual chargers' capacities. As solar is intermittent throughout the cloudy day, the BESS and PEV are not being charged up to their 100 percent capacity. Only a few discharging events take place during on-peak periods.

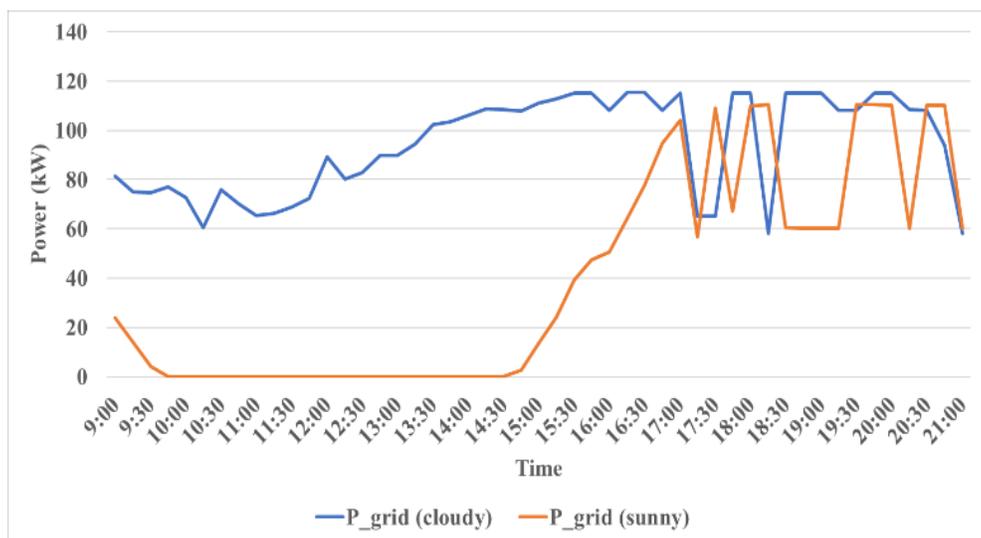

Fig 9: Cloudy Day & Level II Activity: Power Purchased from Grid

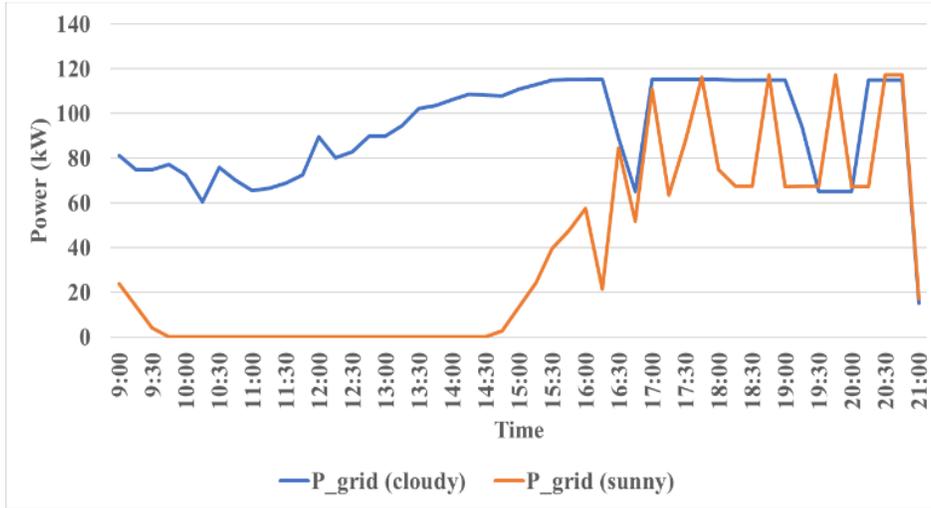

Fig 10: Cloudy Day & Level III Activity: Power Purchased from Grid

*4.7. Impacts on Distribution Feeder*

To analyze the DER impacts on the low voltage distribution grid, IEEE 13 bus system is chosen and modified to adopt the coordinated DERs integrated into the system. This is a relatively highly loaded 4.16 kV feeder equipped with spot loads [62]. To model the test feeder, PV and BESS are connected to node no 671 on phases A and C, respectively. PEV is connected to node 611. HVAC, lighting, and miscellaneous loads are considered to be integrated on node 675 where grid purchased power is added. The nodes are chosen proportionately with maximum power capabilities of different DERs to investigate the real impacts on the feeder. All DERs and building loads are added as additional loads to their respective nodes along with the base spot load values. The modified test feeder is shown in figure 11.

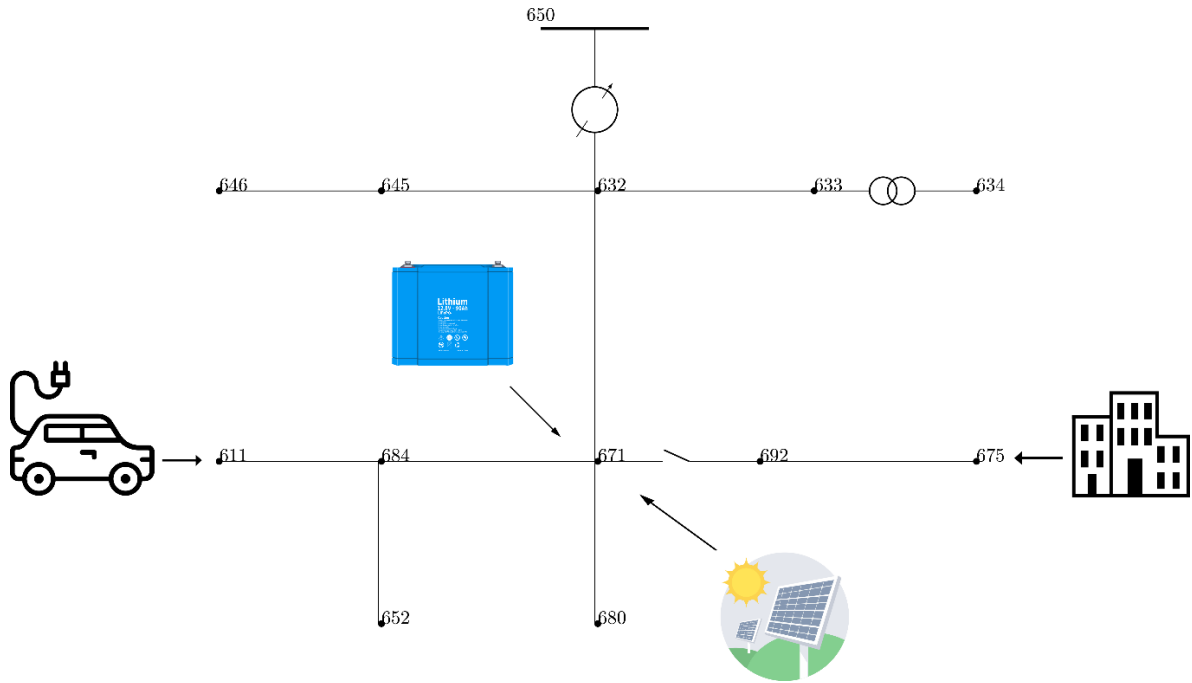

Fig 11: Modified IEEE-13 Bus Test Feeder

In distribution systems, the addition of DERs can make the grid unstable because of their intermittent nature and unoptimized scheduling. To evaluate the DER impacts on a distribution level test feeder, voltage profile study at the respective DER buses is a reliable indicator. Hence, the voltage profiles are analyzed at nodes 611, 671, and 675. To make the BESS, PEV, and PV impacts on the distribution grid explicit, they have been integrated as distributed control equipment instead of connected to the same node. Moreover, if all the DERs are connected to node 675 along with grid-purchased power, it might cause overloading. OpenDSS is used to solve the power flow equation and evaluate the DER impacts on the test feeder. Newton method is used to solve the power flow equation in OpenDSS [63].

While, voltage changes with time help to compare the voltage deviation scenario for any distinct time instant, Voltage Deviation Index (VDI) is highly recommended to show the overall voltage unbalance scenario for the whole simulation period. VDI is formulated as the root mean square voltage deviation from the nominal voltage magnitudes of the buses with time depicted in equation (23). $V_t$ is the per unit (pu) voltage at time t for any node. The nominal voltage is considered as 1 pu. Hence, the voltage deviation is monitored for each time slot in the nodes where DERs are applied and used for VDI calculation.

$$\text{VDI} = \sqrt{\sum_{t=1}^{T}(V_t - V_{nominal})^2 / T} \quad (23)$$

Table 5 shows the VDI for all case scenarios at 611, 671, and 675 nodes. Level III ensures fewer voltage deviations for the building, BESS, and PEV nodes whereas level II results in less VDI for the PV node. Level II activities result in 18% higher VDI for PEV and 13% higher VDI for building nodes respectively. On the other hand, level III results in 24% higher VDI at the PV node.

**Table 5**

Voltage Deviation Index

| Cost Objective | PEV level | Building | PV | BESS | PEV |
|---|---|---|---|---|---|
| Energy+BESS degradation +PEV degradation | Level II | 0.03392 | 0.0279 | 0.07035 | 0.0777 |
| | Level III | 0.0296 | 0.0357 | 0.07 | 0.0644 |

*4.8 Computation Time*

The computational time is a factor in solving any real-time optimization problem. The mathematical modeling done here provides all linear equations to analyze the characteristics of controllable components such as BESS, PEV, HVAC. The linearity provides the flexibility for the solver to solve it quickly and makes it practical for real-time optimization. It takes an average of 0.05 seconds to solve the problems in most of the cases which is a lot faster than any other type of machine learning-based non-linear model.

## 5. Conclusion & Future works

The dynamic cost optimization problem for the customers behind the meter in a modern building-integrated microgrid involves a lot of decisive factors. This paper presents a comprehensive solution incorporating the usual loads like HVAC, lighting and, plug-in loads, with newer technologies like PV, BESS, and PEV. The contributions and limitations of each of these components are represented along with battery degradation in the cost function for optimization. The results show that for buildings equipped with DERs, between 20.5% to 23.0% system cost reduction is possible depending on the type of vehicle chargers. This is valid under the Time of Use (TOU) rate schedule for the most active 12-hour period of a week-day when both the kW demand and kWh energy costs are the highest. For the slower rate of G2V/V2G activities, the maximum increase in kW peak is 39.98 percent for 40 to 60 percent variation in lighting. While for fast G2V/V2G activities, the peak demand increment is 43.07 percent. Effects of electricity price variation are explored which show that operation of BESS or PEV might not be feasible if their degradation costs are higher than the energy costs. This exhibits the need for subsidization or restructuring the utility prices in order to promote these technologies. Effects of weather conditions such as temperature variations and cloudy day impacts are also presented that can be useful in changing the optimization strategy accordingly. The impacts on the distribution grid are depicted for each of the charger types. Contrary to popular belief, it shows that level III charging causes less voltage deviation than level II, which might encourage higher adoption of level III EV charging infrastructures resulting in higher range EV purchases. The influx of heavy-duty electric vehicles and the extreme fast charging capabilities can be explored as a future extension of this work. Moreover, all the stakeholders will be benefitted from a comparative cost-benefit analysis between heavy-duty and light-duty EV integration to the grid.

**Declaration of interest statement**: The authors declare that there are no conflicts of interest.